\documentclass[12pt,preprint,apj]{emulateapj}

\newcommand{\kms}{{\rm km~s^{-1}}}

\newcommand{\ha}{W_{\rm{H} \alpha}}

\shorttitle{AGN fraction of compact group galaxies}
\shortauthors{Sohn et al.}

\begin{document}

\title{Activity in Galactic Nuclei of Compact Group Galaxies in the Local Universe}

\author{Jubee Sohn$^{1}$, Ho Seong Hwang$^{2}$, Myung Gyoon Lee$^{1}$, Gwang-Ho Lee$^{1}$, Jong Chul Lee$^{3}$}

\affil{$^{1}$ Astronomy Program, Department of Physics and Astronomy, 
  Seoul National University, Gwanak-gu, Seoul 151-742, Republic of Korea}
\email{jbsohn@astro.snu.ac.kr}

\affil{$^{2}$ Smithsonian Astrophysical Observatory, 60 Garden Street, Cambridge, MA 02138, USA}
\email{hhwang@cfa.harvard.edu}

\email{mglee@astro.snu.ac.kr}

\email{ghlee@astro.snu.ac.kr}

\affil{$^{3}$ Korea Astronomy and Space Science Institute 776, Daedeokdae-ro, Yuseong-gu, Daejeon 305-348, Republic of Korea}
\email{jclee@kasi.re.kr}


\begin{abstract}
We study the nuclear activity of galaxies in local compact groups.
We use a spectroscopic sample of 238 galaxies in 58 compact groups
  from the Sloan Digital Sky Survey data release 7
  to estimate the fraction of AGN-host galaxies in compact groups,
  and to compare it with those in cluster and field regions.
We use emission-line ratio diagrams to identify AGN-host galaxies,
  and find that the AGN fraction of compact group galaxies is 17-42\% 
  depending on the AGN classification method.
The AGN fraction in compact groups is not the highest
  among the galaxy environments.
This trend remains even if we use several subsamples
  segregated by galaxy morphology and optical luminosity.
The AGN fraction for early-type galaxies decreases 
  with increasing galaxy number density, 
  but the fraction for late-type galaxies changes little.
We find no mid-infrared detected AGN-host galaxies 
  in our sample of compact groups using Wide-field Infrared Survey Explorer data.
These results suggest that the nuclear activity of compact group galaxies 
  (mostly early types) is not strong because of lack of gas supply
  even though they may experience frequent 
  galaxy-galaxy interactions and mergers that could trigger nuclear activity.
\end{abstract}


\section{INTRODUCTION}

Understanding what powers the activity in galactic nuclei
  is one of the key issues
  in the study of galaxy formation and evolution.
There is growing evidence that most galaxies harbor 
  supermassive black holes (SMBHs) in their centers \citep{Kor04,Ho08,McC13}. 
However, it is poorly understood 
  why some galaxies show strong nuclear activity 
  through active mass accretion to SMBHs, but other galaxies do not.  

Several mechanisms for triggering nuclear activity in galaxies
  are suggested to explain a gas inflow 
  towards the center of galaxies to feed SMBHs:
  galaxy-galaxy interaction (e.g., \citealp{San88,Bar92,Spr05,DiM07}), 
  bar-driven gas inflow (e.g., \citealp{Com03}), and stellar wind (e.g., \citealp{Cio07}).
Among these, the galaxy-galaxy interaction has been extensively studied
  because it is expected in the hierarchical picture of galaxy formation
  as the star formation is triggered through galaxy-galaxy interactions and mergers.
For example, 
  numerical simulations showed that both circumnuclear starburst and nuclear activity 
  can be triggered by gas inflow 
  during galaxy-galaxy interactions \citep{Mih96,Spr05,DiM07}.
Some observational studies detected an enhancement of 
  active galactic nuclei (AGN) fraction 
  in galaxy pairs compared to
  isolated galaxies, suggesting a strong connection 
  between galaxy interaction and nuclear activity \citep{Alo07,Ell11}. 
However, there are also other studies that found no significant excess of 
  AGN fraction in galaxy pairs \citep{Sch01,Gro05,Col06,Li08}. 

The studies on the environmental dependence of nuclear activity 
  can provide important hints of AGN triggering mechanism, 
  because the physical mechanisms for nuclear activity, 
  especially galaxy-galaxy interaction, 
  are strongly affected by galaxy environments.
In this regard, several studies investigated 
  the fraction of AGN in field and clusters. 
For example, Sabater et al. (2008) identified AGN candidates among 
  their isolated galaxy sample using far infra-red colors 
  based on IRAS 25 $\mu$m and 60 $\mu$m fluxes,
  and found that 7-22\% of the isolated galaxies host AGN (see also \citealp{Alo07}).
In contrast, other studies suggested a much lower AGN fraction for galaxy clusters. 
Dressler et al. (1999) reported that only 1\% of cluster galaxies 
  show AGN features in the optical spectra, and 
  \citet{Mar07} found that the AGN fraction in galaxy clusters 
  based on X-ray data can increase up to 5\%.
Thus the AGN fraction for high-density regions 
  appears to be smaller than for low-density regions. 
However, the AGN fraction strongly depends on the AGN selection criteria 
  so that this conclusion needs to be checked.
 
Hence, it is very important to apply 
  uniform AGN selection criteria for 
  a homogeneous set of data to understand 
  better environmental dependence of the AGN fraction. 
\citet{Hag10} used uniform AGN selection criteria based on X-ray luminosity from {\it Chandra}
  for field and cluster galaxies in the Sloan Digital Sky Survey (SDSS), 
  and found similar AGN fractions between the two regions, 
  in contrast to the previous results described above.

Recently \citet{Hwa12a} also used uniform AGN selection criteria 
  based on optical spectra for a large sample of 
  cluster and field galaxies in the SDSS. 
They confirmed the difference in the AGN fraction 
  between cluster and field regions 
  in the case of early-type galaxies in the sense that 
  the cluster galaxies show three times higher AGN fraction than the field galaxies. 
However, they found little difference in the AGN fraction 
  between the cluster and field galaxies in the case of late-type galaxies.
From these results they further suggested that the environmental effects on the AGN fraction strongly depend on the host galaxy morphology as well as environments.

Compact groups of galaxies also provide an interesting environment
  to examine the physical mechanisms for nuclear activity.
These groups are isolated association of several galaxies 
  within the compact angular configuration. 
Because of their high galaxy number density (e.g., \citealp{Rub91}) and 
  low velocity dispersions
  ($50-400$ $\kms$ with a median of $\sim$266 $\kms$ \citep{McC09},
  smaller than for rich clusters with $500-1000$ $\kms$ \citep{Rin06}),
  the compact groups are expected to have frequent galaxy-galaxy interactions,
  which could trigger the activity in galactic nuclei.
A formal definition of compact groups was 
  first introduced by \citet{Hic82}:
  $N\geq4$, $\theta_{\rm N}\geq 3 \theta_{\rm G}$, and $\mu_{\rm G}<26.0$ mag arcsec$^{-2}$.
  $N$ is the total number of galaxies within 3 mag of the brightest galaxy.
  $\mu_{\rm G}$ is the mean surface brightness of these galaxies
  within the smallest circular area of $\theta_{\rm G}$, 
  where $\theta_{\rm G}$ is the angular diameter of the smallest circle
  that contains their geometric centers.
  $\theta_{\rm N}$ is the angular diameter of the largest concentric circle
  that contains no additional galaxies within this magnitude range or brighter.
Using these (or slightly modified) selection criteria for compact groups,
  several authors construct compact group catalogs (e.g., \citealp{Hic82,Lee04,McC09}),
  and study the properties of these groups 
  (e.g., \citealp{Hic92,Hic97,Bit10,Bit11,Men11,Coe12}).

In particular, several studies examined how special the compact groups are
  for nuclear activity in galaxies \citep{Coz98,Coz00,Shi00,Gal08,Mar10,Sab12}.
For example, \citet{Coz00} estimated the AGN fraction (41\%) 
  for 193 galaxies in 49 compact groups based on the analysis of the optical spectra.
They suggested that compact groups could be the best location 
  to find AGN in the local universe.
Using the 2MASS and {\it Spitzer} data,
  \citet{Gal08} found that 54\% of galaxies in 12 Hickson compact groups
  show hot dust emissions in the mid-infrared,
  indicating the existence of ongoing nuclear activity and/or star formation.
\citet{Mar10} compiled the optical spectra of 270 galaxies 
  in 64 Hickson compact groups, 
  and suggested that the AGN fraction could be up to 42\% 
  if they include composite galaxies (harboring both AGN and star formation)
  in the AGN class. 
These AGN fractions in compact groups
  apparently are larger than those in field and cluster regions.
However, because of different AGN selection criteria and
  inhomogeneous sample of galaxies in different studies,
  it is very difficult to have a fair comparison
  of the AGN fractions in various environments.
Recently, \citet{Sab12} compared the AGN population of Hickson compact groups 
  and isolated galaxies using a uniform AGN selection criteria based on the optical spectra. 
They found that AGN do not preferentially appear in 
  compact groups after correcting the morphology and luminosity effects. 

In this paper, we study the activity in galactic nuclei of 
  compact group galaxies in the local universe.
We use uniform AGN selection criteria for a homogeneous sample of galaxies 
  to compare nuclear activity of compact group galaxies
  with those of field and cluster galaxies.
This paper is organized as follows.
In Section 2, we describe the sample of compact groups
 and control samples of field and clusters.
Section 3 explains the AGN selection methods.
We present the results on the AGN fraction in compact groups, 
  and compare it with those in other environments in Section 4.
We discuss our results in Section 5 and summarize our main results in Section 6.
Throughout, we adopt flat $\Lambda$CDM cosmological parameters:
  $H_{0} = 70 ~{\rm km ~s^{-1} ~Mpc^{-1}}$, $\Omega_{m} = 0.3$ 
  and $\Omega_{\Lambda} = 0.7$.

\section{DATA}

We use the compact group catalog in \citet[][hereafter M09]{McC09}.
M09 constructed a compact group catalog using 
the SDSS \citep{Yor00} data release 6.
This catalog contains 2297 CGs with 9713 member galaxies.
M09 identified compact groups based only on photometric information. 
Therefore, their sample could be contaminated by interlopers.
They suggested that at least 55\% of their groups may contain interlopers.

To select genuine compact groups without any interlopers,
  we use only compact group galaxies with spectroscopic redshifts in the M09 sample.
We first select 58 compact groups at $0.03 \leq z \leq 0.15$ 
  that contain at least four member galaxies with measured redshifts
  in the SDSS DR7\footnote{We use DR7 to add more spectroscopic redshifts (if available)
  for compact group galaxies originally drawn from the photometric data in DR6.}.
We then select 238 member galaxies with concordant redshifts 
  in these compact groups ($\Delta cz \leq 1000 ~\kms$),
  which will be used for the following analysis.
We emphasize that our results are based only on the spectroscopic sample 
  of compact group galaxies,
  more robust than the results based on the photometric sample of galaxies.
  
We use several value-added galaxy catalogs (VAGCs) 
  for physical parameters of the group galaxies.
Photometric and spectroscopic parameters are adopted 
  from the SDSS pipeline \citep{Sto02} and from the MPA/JHU DR7 
  VAGC\footnote{http://www.mpa-garching.mpg.de/SDSS/DR7/}, respectively.
Morphology information is adopted from 
  the Korea Institute for Advanced Study (KIAS) DR7 
  VAGC\footnote{http://astro.kias.re.kr/vagc/dr7/}\citep{Choi10}.

We use [O {\scriptsize III}] line fluxes taken from the MPA/JHU VAGC \citep{Tre04},
  which is the straight integration over the fixed bandpass from the
  continuum-subtracted emission line.
We correct internal extinction of the line fluxes 
  using the Balmer decrement and reddening curve given by \citet{Car89}
  by assuming an intrinsic H$\alpha$/H$\beta$ value of 3.1 for AGN host-galaxies.
We compute black hole masses ($M_{\rm BH}$) for AGN-host galaxies
  using the $M_{\rm BH}-\sigma$ relation:
  ${\rm log} (M_{\rm BH}/M_\odot) = \alpha + \beta {\rm log}(\sigma/200$ km s$^{-1}$).
We adopt
  $\alpha=8.39, \beta=5.20$ for early-type galaxies,
  $\alpha=8.07, \beta=5.06$ for late-type galaxies, respectively \citep{McC13}.
 
\begin{figure}
\centering
\includegraphics[scale=0.9]{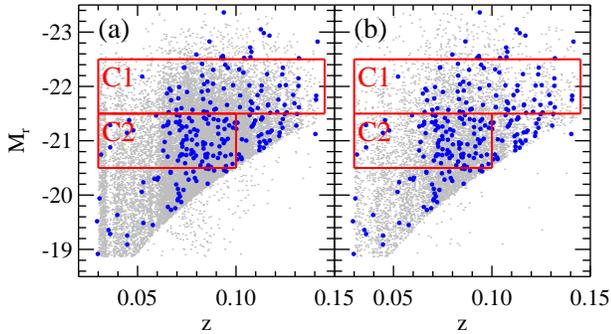}
\caption{$M_r$ -- $z$ diagrams for the galaxies in the compact groups in our sample (filled circles), field (dots in (a)), and clusters (dots in (b)) in the SDSS. 
Larger boxes represent boundaries for the two magnitude subsamples.}
\label{control}
\end{figure}

We show $r$-band absolute magnitudes of our compact group galaxies
  as a function of redshifts in Figure \ref{control} (blue filled circles in both panels).
For comparison, 
  we also plot the galaxies in the field (gray dots in the left panel) and 
  in galaxy clusters (gray dots in the right panel).
We use these field and cluster galaxies to 
  compare their nuclear activity with those of compact group galaxies
   in Section \ref{compenv}.
The field and cluster galaxy samples are taken from \citet{Hwa12a}
  who use the same spectroscopic sample of galaxies in SDSS DR7.
They compiled galaxies with spectroscopic redshifts 
  associated with 129 relaxed Abell clusters at $0.02 < z < 0.14$.
Among them, 
  we use 7211 cluster galaxies within the virial radii of clusters ($r_{\rm 200,cl}$),
  and 20,331 field galaxies outside the cluster region 
  (R $> 5 r_{\rm 200,cl}$).
$R$ is the projected clustercentric radius.
To study the environmental dependence of 
  the AGN fraction at a fixed optical luminosity, 
  we divide the galaxies into two magnitude subsamples 
  based on absolute $r$-band magnitudes and redshifts:
  a bright subsample 
  (C1; $-22.5 \leq M_{r} < -21.5$ and $0.030 \leq z \leq 0.145$)
  and a faint subsample 
  (C2; $-21.5 \leq M_{r} < -20.5$ and $0.030 \leq z \leq 0.100$). 
The C1 and C2 samples contain 64 and 83 galaxies, respectively. 

\section{AGN SELECTION}

To identify AGN-host galaxies in our sample,
  we use three methods based on emission-line flux ratios:
  one with strong emission lines and two with weak emission lines.

\subsection{Strong emission-line galaxy classification}

We first use the Baldwin-Phillips-Terlevich (BPT) line ratio diagrams 
  based on 
  [O {\scriptsize III}]/H$\beta$ and [N {\scriptsize II}]/H$\alpha$ \citep{BPT81}.
Among 238 compact group galaxies,
  we apply this method only to 83 galaxies
  with signal-to-noise ratio (S/N) $\geq 3$ in the emission lines 
  H$\alpha$, H$\beta$, [O {\scriptsize III}] $\lambda5007$, 
  and [N {\scriptsize II}] $\lambda6584$.
We show the line ratios of these 83 galaxies
  in the left panel in Figure \ref{class}.
We classify them (star-forming galaxies, AGN, and composite galaxies) 
  based on their relative positions with respect to the demarcation lines
  identifying extreme starbursts \citep{Kew01} and pure SF \citep{Kau03}. 

Composite galaxies are between the two demarcation lines. 
They could be extreme star-forming galaxies \citep{Kew01}, 
  or host a mixture of star formation and AGN \citep{Kew06}. 
Although the physical origin of this class is not completely understood \citep{Ho08},
  several observational results in other wavelengths suggest 
  that many of them host (hidden) AGN (e.g., \citealp{Pan05,Lee12}). 
In this study, 
  we include composite galaxies in the AGN class 
  to compute the AGN fraction. 
We call ‘pure SF’ and ‘pure AGN’ for star-forming and AGN types 
  (pSF and pAGN in tables and figures), respectively, 
  to distinguish them from composite galaxies. 
Excluding composite galaxies to compute the AGN fraction 
  does not change our conclusions, although our statistics become worse 
  as the number of AGN-host galaxies becomes smaller.

\subsection{Weak emission-line galaxy classification}

To determine the optical spectral types 
  of the remaining 155 compact group galaxies with low-S/N emission lines,
  we use two methods based on flux ratios of weak emission lines.
  
\begin{figure*}
\centering
\includegraphics[scale=0.9]{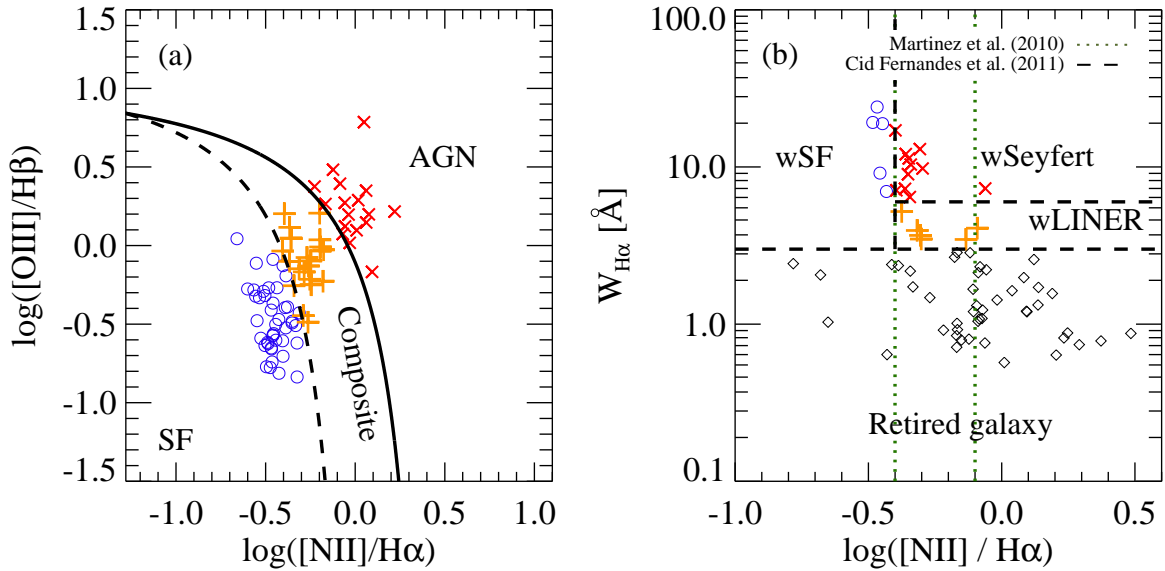}
\caption{Classification of spectral types for compact group galaxies.
(a) The BPT diagnostic diagram with [O {\scriptsize III}]/H$\beta$ vs. [N {\scriptsize II}]/H$\alpha$ emission-line ratios.
The dashed line represents a boundary for pure star-forming galaxies \citep{Kau03} and the solid line denotes a theoretical upper limit for starburst galaxies provided by \citet{Kew01}.
(b) The WHAN diagram \citep{Cid11} for weak emission-line galaxies with ${\rm S/N}<3$ of H$\beta$ or [O {\scriptsize III}].
Dashed lines are borderlines that divide weak emission-line galaxies into weak emission-line SFs (wSF), weak emission-line Seyferts (wSeyfert), weak emission-line LINER (wLINER), and retired galaxies.
Two vertical dotted lines represent the classification scheme for star-forming nuclei, transition objects, and AGN  in \citet{Mar10}.}
\label{class}
\end{figure*}

We use the WHAN method \citep{Cid10,Cid11} that
  uses [N {\scriptsize II}]/H$\alpha$ and H$\alpha$ equivalent width ($\ha$).
This method classifies galaxies into five spectral types:
  pure SFs, Seyferts, LINERs, retired galaxies
  and passive galaxies (no emission lines). 
We can apply this method only to 71 galaxies with
  S/N$\geq$3 in [N {\scriptsize II}] and H$\alpha$ emission lines.
We show the classification diagram 
  in the right panel of Figure \ref{class} (dashed lines),
  and list the selection criteria for each spectral type,
\begin{itemize}
\item weak emission-line star-forming galaxies: log([N {\scriptsize II}]/H$\alpha) > -0.4$ and $\ha \geq 3$ \AA;
\item weak emission-line Seyferts: log([N {\scriptsize II}]/H$\alpha) > -0.4$ and $\ha \geq 6$ \AA;
\item weak emission-line LINERs: log([N {\scriptsize II}]/H$\alpha) > -0.4$ and 3 \AA $~\leq \ha < 6$ \AA.
\end{itemize}

We include weak emission-line Seyferts and LINERs in the AGN class 
  when we compute the AGN fraction using this method.

We also adopt the classification method in \citet{Mar10}, 
  which is based only on [N {\scriptsize II}]/H$\alpha$ (hereafter N2H$\alpha$ method). 
Their classification scheme is as follows 
  (vertical dotted lines in the right panel of Figure \ref{class}):
\begin{itemize}
\item star-forming galaxies: log([N {\scriptsize II}]/H$\alpha) \leq -0.4$;
\item composite galaxies: $-0.4 <$ log([N {\scriptsize II}]/H$\alpha) \leq -0.1$;
\item AGN: log([N {\scriptsize II}]/H$\alpha) > -0.1$.
\end{itemize}

\section{RESULTS}

\subsection{AGN fraction in compact groups}

\begin{deluxetable*}{lccccccccccccc}
\tabletypesize{\scriptsize}
\tablecolumns{14}
\tablewidth{0pt}
\setlength{\tabcolsep}{0.1in}
\tablecaption{Spectral Types of All Galaxies in Compact Groups}
\tablehead{
\colhead{} & \colhead{} & & \multicolumn{2}{c}{pSF} & & \multicolumn{2}{c}{Composite}
& & \multicolumn{2}{c}{pAGN\tablenotemark{d}} & & \multicolumn{2}{c}{AGN\tablenotemark{e}}\\
\cline{4-5}
\cline{7-8}
\cline{10-11}
\cline{13-14}
Method & N(sample)\tablenotemark{a} &	& \colhead{N\tablenotemark{b}} & \colhead{$f_{\rm{pSF}} (\%)$\tablenotemark{c}}
& & \colhead{N} & \colhead{$f_{\rm{Comp}} (\%)$} & & \colhead{N} & \colhead{$f_{\rm{pAGN}} (\%)$}
& & \colhead{N} & \colhead{$f_{\rm{AGN}} (\%)$}}
\startdata
BPT 			  &	 83	& & 43	& $18.1\pm2.5$	& &	23	& $9.7\pm1.9$	& &	17	& $7.1\pm1.6$  & &	40	& $16.8\pm2.5$\\
WHAN              &  71 & & 5   & $2.1\pm0.9$   & & --	& --     & & 18  & $7.6\pm1.7$  & &	18	& $7.6\pm1.7$\\
N2H$\alpha$       &  71 & & 10	& $4.2\pm1.2$	& &	31	& $13.0\pm2.1$	& &	30	& $12.6\pm2.2$ & &	61	& $25.6\pm2.7$\\
BPT + WHAN 		  &	154	& & 48	& $20.2\pm2.6$	& & --	& --		& & 58	& $24.4\pm2.8$ & &	58	& $24.4\pm2.8$\\
BPT + N2H$\alpha$ &	154	& & 53	& $22.3\pm2.7$	& & 54	& $22.7\pm2.8$	& & 47	& $19.7\pm2.5$ & &	101	& $42.4\pm3.3$
\enddata
\tablenotetext{a}{The number of galaxies used for spectral classification.}
\tablenotetext{b}{The number of galaxies in each spectral type.}
\tablenotetext{c}{The fraction of each type among total 238 compact group galaxies in our sample. 
The error in the fraction represents 68\% ($1\sigma$) confidence interval 
  that is determined from the bootstrap resampling method.}
\tablenotetext{d}{Pure AGN including Seyferts and LINERs.}
\tablenotetext{e}{AGN including pAGN and composite galaxies.}
\label{cg_agn}
\end{deluxetable*}

Among the 238 compact group galaxies in our sample,
  we can determine the spectral types of 83 galaxies with high-S/N emission lines
  based on the BPT method: 
  43 pure SFs, 23 composite galaxies, and 17 pure AGN.
These correspond to $51.8\% \pm 5.4\%$, $27.7\% \pm 5.0\%$, 
  and $20.5\% \pm 4.4\%$ among the strong emission-line galaxies.
If we compute the fraction 
  among the total sample of compact group galaxies, 
  the fraction of each spectral type would be 
  $18.1\% \pm 2.5\%$, 
   $9.7\% \pm 1.9\%$, and 
   $7.1\% \pm 1.6\%$
   for pure star-forming, composite, and pure AGN-host galaxies, respectively.
The AGN fraction (including composite galaxies) 
  is then $16.8\% \pm 2.5\%$.
We summarize these results in Table \ref{cg_agn}.
We also list the results for 71 weak emssion-line galaxies 
  based on WHAN and N2H$\alpha$ methods in Table \ref{cg_agn}.

Table \ref{cg_agn} indicates that
  the AGN fractions based on strong plus weak emission-line galaxies
  are higher than for based only on strong emission-line galaxies. 
If we combine BPT and WHAN methods, 
  the total AGN fraction is $24.4\% \pm 2.8\%$.
On the other hand, 
  the total AGN fraction is $42.4\% \pm 3.3\%$
  if we combine BPT and N2H$\alpha$ methods.
We discuss in Section \ref{compother}
  why different AGN selection methods give different AGN fractions.
  
We also examine the AGN fractions of compact group galaxies
  segregated by their morphologies.
Among the 238 compact group galaxies,
  there are 152 (64\%) early-type and 86 (36\%) late-type galaxies.
The higher fraction of early-type galaxies in compact groups
  is also apparent for the compact groups in \citet{McC09} and 
  for the Hickson compact groups \citep{Hic82}.
The spectra of early-type galaxies typically do not show emission lines.
Therefore, we can determine the spectral types based on the BPT method
  only for 17.8\% (27 galaxies) of early-type galaxies.
Using the WHAN and the N2H$\alpha$ methods,
  we can increase the number of early-type galaxies with spectral types
  up to 73 (48.0\%).
For late-type galaxies,
  we can determine the spectral types 
  for 65.1\% (56 galaxies) based on the BPT method,
  and 94.2\% (81 galaxies)
  based on the combination of BPT and N2H$\alpha$/WHAN methods.

We summarize the fraction of each spectral type 
  segregated by galaxy morphologies in Table \ref{morph_class}.
As expected, the AGN fractions are different depending on the
  AGN selection method.
However, all the AGN selection methods yield
  higher AGN fractions for late-type galaxies than for early-type galaxies.

\begin{deluxetable*}{lcccccccccccccc}
\tabletypesize{\scriptsize}
\tablecolumns{15}
\tablewidth{0pt}
\setlength{\tabcolsep}{0.1in}
\renewcommand{\arraystretch}{1.2}
\tablecaption{Spectral Types of Early- and Late-type Galaxies in Compact Groups}
\tablehead{
 &  &  & & \multicolumn{2}{c}{pSF} & & \multicolumn{2}{c}{Composite}
& & \multicolumn{2}{c}{pAGN} & & \multicolumn{2}{c}{AGN}\\
\cline{5-6}
\cline{8-9}
\cline{11-12}
\cline{14-15}
\colhead{Method}	& \colhead{Types} & \colhead{N(sample)} & & \colhead{N} & \colhead{$f_{\rm{pSF}} (\%)$\tablenotemark{a}} & & \colhead{N} & \colhead{$f_{\rm{Comp}} (\%)$} & & \colhead{N} & \colhead{$f_{\rm{pAGN}} (\%)$}
& & \colhead{N} & \colhead{$f_{\rm{AGN}} (\%)$}}
\startdata
BPT			  & ETG & 27 & & 9  &  $5.9\pm1.8$ 	& &	8	& $5.3\pm1.7$	& &	10	& $6.6\pm2.0$ & &	18	& $11.8\pm2.6$ \\
			  & LTG & 56 & & 34 & $39.5\pm5.2$	& &	15	& $17.4\pm4.1$	& &	7	& $8.1\pm2.9$ & &	22	& $25.6\pm4.7$ \\
\cline{1-15}
BPT 		  & ETG & 73 & & 9  &  $5.9\pm1.8$	& & --  & --		& & 23	& $15.1\pm2.8$ & &	23	& $15.1\pm2.8$ \\
+ WHAN		  & LTG & 81 & & 39 & $45.3\pm5.5$	& & --	& --		& & 34	& $39.5\pm5.2$ & &	34	& $39.5\pm5.2$ \\
\cline{1-15}				
BPT 		  & ETG & 73 & & 14 & $9.2\pm2.3$ 	& &	25	& $16.4\pm2.9$	& &	34	& $22.4\pm3.3$ & &	59	& $38.8\pm4.0$  \\
+ N2H$\alpha$ & LTG & 81 & & 39 & $45.3\pm5.3$	& &	29	& $33.7\pm5.0$	& &	13	& $15.1\pm3.8$ & &	42	& $48.8\pm5.4$
\enddata
\tablenotetext{a}{The fraction whose numerator is the number of ETGs (or LTGs) with a spectral type
and whose denominator is the number of total 152 ETGs (or total 86 LTGs).}
\label{morph_class}
\end{deluxetable*}

\subsection{Comparison of AGN fractions in various environments}

\begin{figure*}
\centering
\includegraphics[scale=1.0]{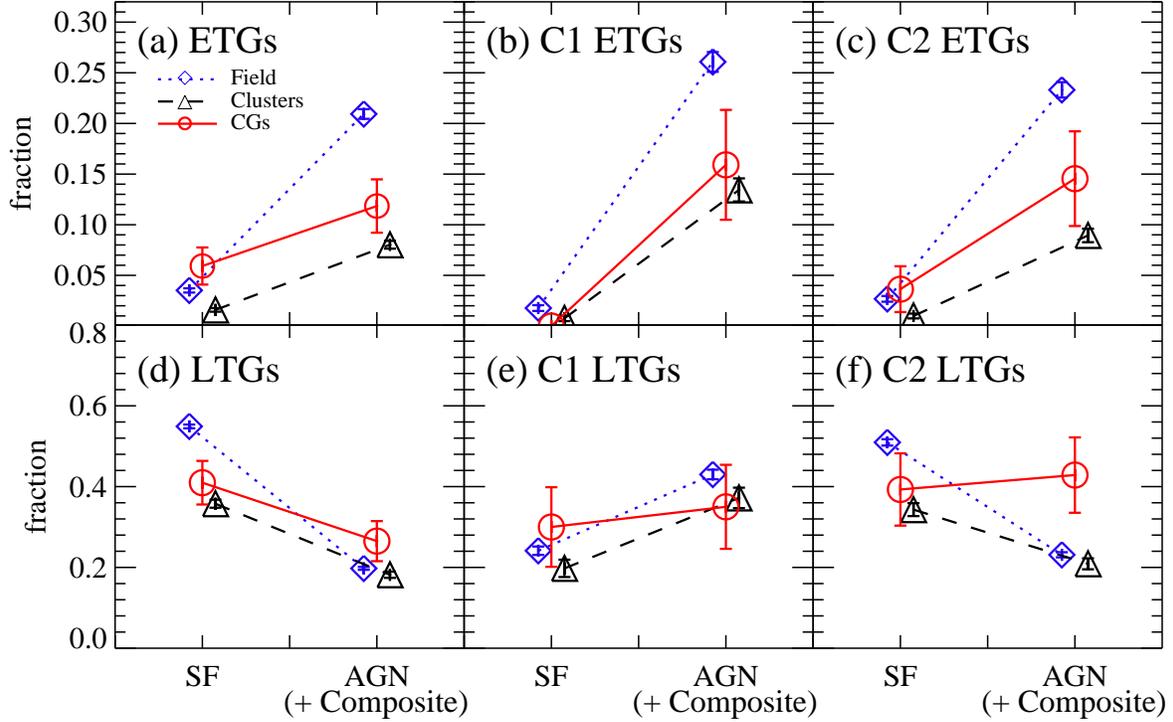}
\caption{Fractions of each spectral type for early-type (upper panels) and late-type (lower panels) galaxies in different galaxy environments (field: dotted-diamonds, clusters: dashed-triangles, and compact groups: solid-circles).
Fractions are for the whole sample (a,d), the C1 sample (b,e), and the C2 sample (c,f).
We arbitrarily shifted the points along the x-axis to show symbols and error bars clearly.}
\label{env}
\end{figure*}

To study the environmental dependence of nuclear activity,
  we plot the fractions of each spectral type for compact group galaxies (red solid line)
  in comparison with cluster (black dashed line) and field (blue dotted line) galaxies
  in Figure \ref{env}.
The fractions are based on the BPT method, and 
  the results do not change even if we use the AGN fractions based on 
  other methods (i.e., WHAN and N2H$\alpha$).
Because the AGN fraction strongly depends on 
  host galaxy properties \citep{Choi09,Sch10},
  we compute the fractions of each spectral type 
  segregated by absolute magnitude and morphology.
The top panels are for early-type galaxies, 
  and bottom panels are for late-type galaxies.
The left, middle and right panels are for total, bright and faint samples of galaxies.

The top left panel shows that
  the AGN fraction (including composite galaxies) 
  for early-type galaxies is not the highest in compact groups. 
This is also apparent for two magnitude subsamples (middle and right panels).
The AGN fraction seems to decrease from field to cluster regions.
The AGN fraction for late-type galaxies 
  is not higher in compact groups compared to other regions (left bottom panels), 
  similar to early-type galaxies.
However, the difference in AGN fractions 
  among different environments is insignificant for late-type galaxies.
The two magnitude subsamples show similar trends.
Although the C2 sample shows the higher AGN fraction in compact groups, 
  the difference is insignificant ($<2\sigma$) due to the large error. 
The fractions of star-forming galaxies among late types are again similar
  in all environments.
We also summarize these results in Table \ref{compare}, 
  \ref{tab_env_egal} and \ref{tab_env_lgal}.
  
\begin{deluxetable*}{lccccccccc}
\tabletypesize{\scriptsize}
\tablecolumns{10}
\tablewidth{0pt}
\setlength{\tabcolsep}{0.1in}
\renewcommand{\arraystretch}{1.3}
\tablecaption{Comparison of Spectral Types of Galaxies in the Field, Compact Groups, and Clusters}
\tablehead{
Environment & N(total) 		& \multicolumn{2}{c}{pSF} & \multicolumn{2}{c}{Composite} & \multicolumn{2}{c}{pAGN}	&	\multicolumn{2}{c}{AGN} \\
\cline{3-4}
\cline{5-6}
\cline{7-8}
\cline{9-10}
\colhead{}	& \colhead{}	& \colhead{N}			  & \colhead{$f_{\rm{pSF}}$ (\%)} & \colhead{N} 					& \colhead{$f_{\rm{Comp}}$ (\%)} & \colhead{N} & \colhead{$f_{\rm{pAGN}}$ (\%)} & \colhead{N} & \colhead{$f_{\rm{AGN}}$ (\%)} } 
\startdata
Field		& 20331 &	6917	& $34.0\pm0.3$	& 2393	&	$11.8\pm0.2$	&  1705	&	$8.4\pm0.2$ & 4098	&	$20.2\pm0.3$ \\
Compact Groups	& 238 	&	43	& $18.1\pm2.5$	& 23	&	$ 9.6\pm2.0$	&  17	&	$7.1\pm1.6$ & 40	&	$16.8\pm2.5$ \\
Clusters	& 7211 	&	911		& $12.6\pm0.4$	& 464	&	$ 6.4\pm0.3$	&  349	&	$4.8\pm0.2$ & 813	&	$11.3\pm0.4$ 
\enddata
\label{compare}
\end{deluxetable*}

\begin{deluxetable*}{lccccccccccccc}
\tabletypesize{\scriptsize}
\tablecolumns{14}
\tablewidth{0pt}
\setlength{\tabcolsep}{0.1in}
\renewcommand{\arraystretch}{1.2}
\tablecaption{Comparison of Spectral Types of Early-type Galaxies in the Field, Compact Groups, and Clusters (BPT)}
\tablehead{
\multicolumn{4}{l}{(1) Early-type galaxies in the total sample} &&&&&&&&& \\
\cline{1-14}
Environment & N(total)\tablenotemark{a} & N(sample)\tablenotemark{b} & \multicolumn{2}{c}{pSF} & & \multicolumn{2}{c}{Composite}
& & \multicolumn{2}{c}{pAGN} & & \multicolumn{2}{c}{AGN}\\
\cline{4-5}
\cline{7-8}
\cline{10-11}
\cline{13-14}
\colhead{}  & \colhead{} & \colhead{}	& \colhead{N} & \colhead{$f_{\rm{pSF}}$ (\%)} & & \colhead{N} & \colhead{$f_{\rm{Comp}}$ (\%)}
& & \colhead{N} & \colhead{$f_{\rm{pAGN}}$ (\%)} & & \colhead{N} & \colhead{$f_{\rm{AGN}}$ (\%)} }
\startdata
Field		& 8185 &	2001	&	288		& $3.5\pm0.2$	& & 753		&	$9.2\pm0.3$		& &	960		&	$11.7\pm0.3$ & &	1713	&	$20.9\pm0.5$ \\
Compact Groups		& 152  &	27	&	9		& $5.9\pm1.8$	& & 8		&	$5.3\pm1.7$		& &	10		&	$6.6\pm2.0$  & &	18		&	$11.8\pm2.6$ \\
Clusters	& 4849 &	466	&	77		& $1.6\pm0.2$	& & 201		&	$4.1\pm0.3$		& &	188		&	$3.9\pm0.3$  & &	389		&	$8.0\pm0.4$ \\
\cline{1-14}

\multicolumn{4}{l}{(2) Early-type galaxies in the C1 sample} &&&&&&&&&\\
\cline{1-14}
Environment & N(total) & N(sample) & \multicolumn{2}{c}{pSF} & & \multicolumn{2}{c}{Composite}
& & \multicolumn{2}{c}{pAGN} & & \multicolumn{2}{c}{AGN} \\
\cline{4-5}
\cline{7-8}
\cline{10-11}
\cline{13-14}
& & & N & $f_{\rm{pSF}} (\%)$ & & N & $f_{\rm{Comp}} (\%)$ & & N & $f_{\rm{pAGN}} (\%)$ & & N & $f_{\rm{AGN}} (\%)$\\
\cline{1-14}
Field		& 2033 &	566	&	36		& $1.8\pm0.3$	& & 173		&	$8.5\pm0.6$		& &	357		&	$17.6\pm0.9$ & &	530		&	$26.1\pm1.0$\\
Compact Groups			& 44   &	7	&	0		& $0.0\pm0.0$	& & 1		&	$2.3\pm1.6$		& &	6		&	$13.6\pm5.0$ & &	7		&	$15.9\pm5.5$\\
Clusters	& 923  &	131	&	7		& $0.8\pm0.3$	& & 58		&	$6.3\pm0.8$		& &	66		&	$7.2\pm0.8$  & &	124		&	$13.4\pm1.1$\\
\cline{1-14}

\multicolumn{4}{l}{(3) Early-type galaxies in the C2 sample} &&&&&&&&&\\
\cline{1-14}
Environment & N(total) &	N(sample) & \multicolumn{2}{c}{pSF} & & \multicolumn{2}{c}{Composite}
& & \multicolumn{2}{c}{pAGN} & & \multicolumn{2}{c}{AGN}\\
\cline{4-5}
\cline{7-8}
\cline{10-11}
\cline{13-14}
& & & N & $f_{\rm{pSF}} (\%)$ & & N & $f_{\rm{Comp}} (\%)$ & & N & $f_{\rm{pAGN}} (\%)$ & & N & $f_{\rm{AGN}} (\%)$ \\
\cline{1-14}
Field		& 3209 &	834	&	86		& $2.7\pm0.3$	& & 330		&	$10.3\pm0.5$	& &	418		&	$13.0\pm0.6$    & &	748		&	$23.3\pm0.7$\\
Compact Groups			& 55   &	10	&	2		& $3.6\pm2.3$	& & 5		&	$9.1\pm3.7$		& &	3		&	$5.5\pm2.9$     & &	8		&	$14.5\pm4.6$\\
Clusters	& 1812 & 	180	&	18	    & $1.0\pm0.2$	& & 85		&	$4.7\pm0.5$		& &	77		&	$4.2\pm0.5$     & &	162		&	$8.9\pm0.7$
\enddata
\tablenotetext{a}{The number of total galaxies in each category.}
\tablenotetext{b}{The number of galaxies classified their spectral type using the BPT method.}
\label{tab_env_egal}
\end{deluxetable*}


\begin{deluxetable*}{lccccccccccccc}
\tabletypesize{\scriptsize}
\tablecolumns{14}
\tablewidth{0pt}
\setlength{\tabcolsep}{0.1in}
\renewcommand{\arraystretch}{1.2}
\tablecaption{Comparison of Spectral Types of Late-type Galaxies in the Field, Compact Groups, and Clusters (BPT)}
\tablehead{
\multicolumn{4}{l}{(1) Late-type galaxies in the total sample} &&&&&&&&&\\
\cline{1-14}
Environment & N(total) & N(sample) & \multicolumn{2}{c}{pSF} & & \multicolumn{2}{c}{Composite} & & \multicolumn{2}{c}{pAGN} & & \multicolumn{2}{c}{AGN}\\
\cline{4-5}
\cline{7-8}
\cline{10-11}
\cline{13-14}
\colhead{}  & \colhead{} & \colhead{} & \colhead{N} & \colhead{$f_{\rm{pSF}}$ (\%)} & & \colhead{N} & \colhead{$f_{\rm{Comp}}$ (\%)}
& & \colhead{N} & \colhead{$f_{\rm{pAGN}}$ (\%)} & & \colhead{N} & \colhead{$f_{\rm{AGN}}$ (\%)} }
\startdata
Field		& 12,071  &	9013	& 6628	& $54.9\pm0.4$	& & 1640	&	$13.6\pm0.3$	& &	745		&	$6.2\pm0.2$ & &	2385	&	$19.8\pm0.4$\\
Compact Groups			& 86      &	56	& 34		& $39.5\pm5.2$	& & 15		&	$17.4\pm4.1$	& &	7		&	$8.1\pm2.9$ & &	22		&	$25.6\pm4.7$\\
Clusters	& 2332    &	1258	& 834		& $35.8\pm1.0$	& & 263		&	$11.3\pm0.6$	& &	161		&	$6.9\pm0.5$ & &	424		&	$18.2\pm0.8$\\
\cline{1-14}

\multicolumn{4}{l}{(2) Late-type galaxies in the C1 sample} &&&&&&&&&\\
\cline{1-14}
Environment & N(total) & N(sample) & \multicolumn{2}{c}{pSF} & & \multicolumn{2}{c}{Composite} & & \multicolumn{2}{c}{pAGN} & & \multicolumn{2}{c}{AGN}\\
\cline{4-5}
\cline{7-8}
\cline{10-11}
\cline{13-14}
& & & N & $f_{\rm{pSF}} (\%)$ & & N & $f_{\rm{Comp}} (\%)$ & & N & $f_{\rm{pAGN}} (\%)$ & & N & $f_{\rm{AGN}} (\%)$\\
\cline{1-14}
Field		& 1717    &	1152	&	414		& $24.1\pm1.0$	& & 408		&	$23.8\pm1.0$	& &	330		&	$19.2\pm1.0$ & &	738		&	$43.0\pm1.2$\\
Compact Groups			& 20      &	13	&	6		& $30.0\pm10.1$	& & 3		&	$15.0\pm7.7$	& &	4		&	$20.0\pm8.3$ & &	7		&	$20.0\pm11.0$\\
Clusters	& 344     &	196	&	68		& $19.8\pm2.1$	& & 60		&	$17.4\pm2.0$	& &	68		&	$19.8\pm2.1$ & &	128		&	$37.2\pm2.7$\\
\cline{1-14}

\multicolumn{4}{l}{(3) Late-type galaxies in the C2 sample} &&&&&&&&&\\
\cline{1-14}
Environment & N(total) & N(sample) & \multicolumn{2}{c}{pSF} & & \multicolumn{2}{c}{Composite} & & \multicolumn{2}{c}{pAGN} & & \multicolumn{2}{c}{AGN}\\
\cline{4-5}
\cline{7-8}
\cline{10-11}
\cline{13-14}
& & & N & $f_{\rm{pSF}} (\%)$ & & N & $f_{\rm{Comp}} (\%)$ & & N & $f_{\rm{pAGN}} (\%)$ & & N & $f_{\rm{AGN}} (\%)$\\
\cline{1-14}
Field		& 4359    &	3229	&	2221	& $51.0\pm0.7$	& & 751		&	$17.2\pm0.6$	& &	257		&	$5.9\pm0.4$ & &	1008	&	$23.1\pm0.6$\\
Compact Groups			& 28      &	23	&	11		& $39.3\pm9.7$	& & 9		&	$32.1\pm8.7$	& &	3		&	$10.7\pm5.6$ & &	12  	&	$42.9\pm9.6$\\
Clusters	& 828     &	457	&	284		& $34.3\pm1.6$	& & 108		&	$13.0\pm1.2$	& &	65		&	$7.9\pm0.9$ & &	173		&	$20.9\pm1.5$
\enddata
\label{tab_env_lgal}
\end{deluxetable*}

To further examine the environmental dependence of nuclear activity, 
  we use a surface galaxy number density ($\Sigma_{3}$) 
  as an environment indicator.
We compute the surface galaxy number density with $\Sigma_{3} = 3(\pi D^{2}_{p,3})^{-1}$, 
  where $D_{p,3}$ is a projected distance to the third nearest galaxy.
To minimize the contamination by foreground and background galaxies, 
  we identify the third nearest galaxies 
  among the neighbor galaxies with M$_{r} < -21.5$ and $0.03 \leq z \leq 0.15$ 
  from the spectroscopic sample of galaxies in the SDSS DR7.
The neighbor galaxies should have relative velocities to the target galaxy 
  smaller than $1500 ~\kms$.

\begin{figure*}
\centering
\includegraphics[scale=0.95]{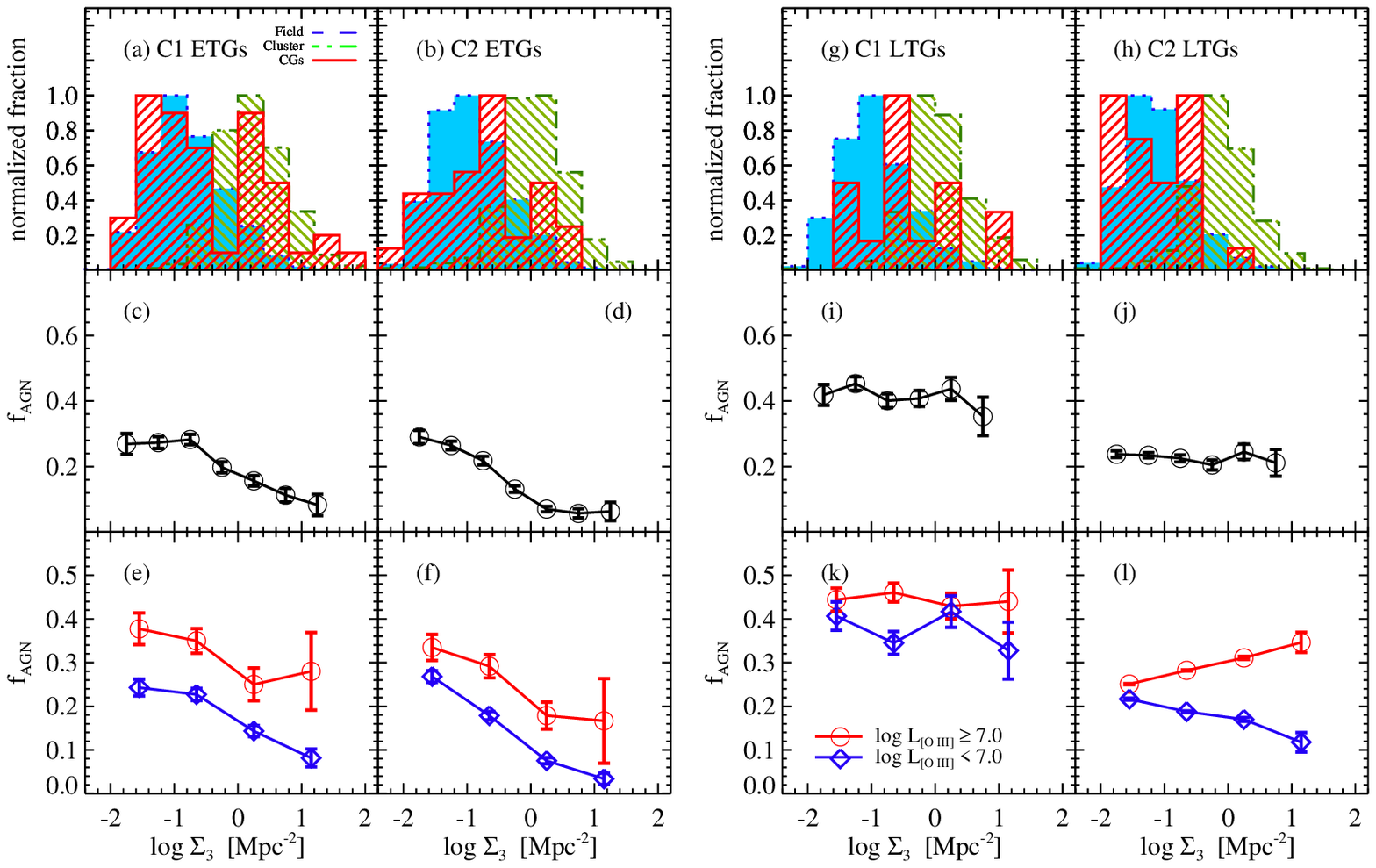}
\caption{Left: AGN fraction as a function of the surface galaxy number density, $\Sigma_{3}$, for (a) bright and (b) faint early-type galaxies.
The normalized fractions of $\Sigma_{3}$ of early-type galaxies in the field (dashed-filled), clusters (dot-dashed-hatched) and compact groups (solid-hatched) for (c) the C1 and (d) C2 subsamples.
The bottom panels show the fraction of strong AGN-host galaxies ($\log L_{[\rm O III]} \geq 7.0$, circles), and weak AGN--host galaxies ($\log L_{[\rm O III]} < 7.0$, diamonds) for each subsample. 
Right: Same as left panels, but for late-type galaxies.}
\label{sig}
\end{figure*}

We show the normalized distribution of $\Sigma_{3}$ in top panels
  of Figure \ref{sig}.  
The left two panels are for early-type galaxies, 
  and right two panels are for late-type galaxies. 
$\Sigma_{3}$ gradually increases from field (dashed-filled histogram) 
  to cluster (dot-dashed-hatched histogram) in all panels. 
Compact group (solid-hatched histogram) galaxies are distributed in wide range, 
  generally in between field and cluster galaxies, 
  except bright early-type galaxies in C1 sample. 

The middle panels show the AGN fractions as a function of $\Sigma_{3}$
  for each subsample.
For early-type galaxies, the AGN fraction in high-density regions 
  is lower than for low-density regions 
  in both two magnitude subsamples (panels (c-d)).
The difference is larger for relatively faint galaxy sample (panel (d)).
On the other hand, 
  the AGN fraction for late-type galaxies does not change with 
  surface galaxy number density (panels (i-j)).
These results are consistent with those in Figure \ref{env}, 
  which are based on different subsamples rather than surface galaxy number densities.
  
In bottom panels,
  we show the fractions of AGN-host galaxies segregated by their 
  [O III] line luminosities.
The [O III] line luminosity could be an accretion rate indicator 
  (\citealt{Kau03,Hec05}; but see also \citealt{tb10}).
The left two panels for early-type galaxies
  show that the AGN fractions again decrease with
  increasing surface galaxy number density
  except the strong AGN ($\log L_{[\rm O III]} \geq 7.0$) in C1 sample.
For late-type galaxies,
  the dependence of AGN fraction for each subsample is not conclusive,
  but no subsamples show strong dependence.  

\begin{figure}
\centering
\includegraphics[scale=0.5]{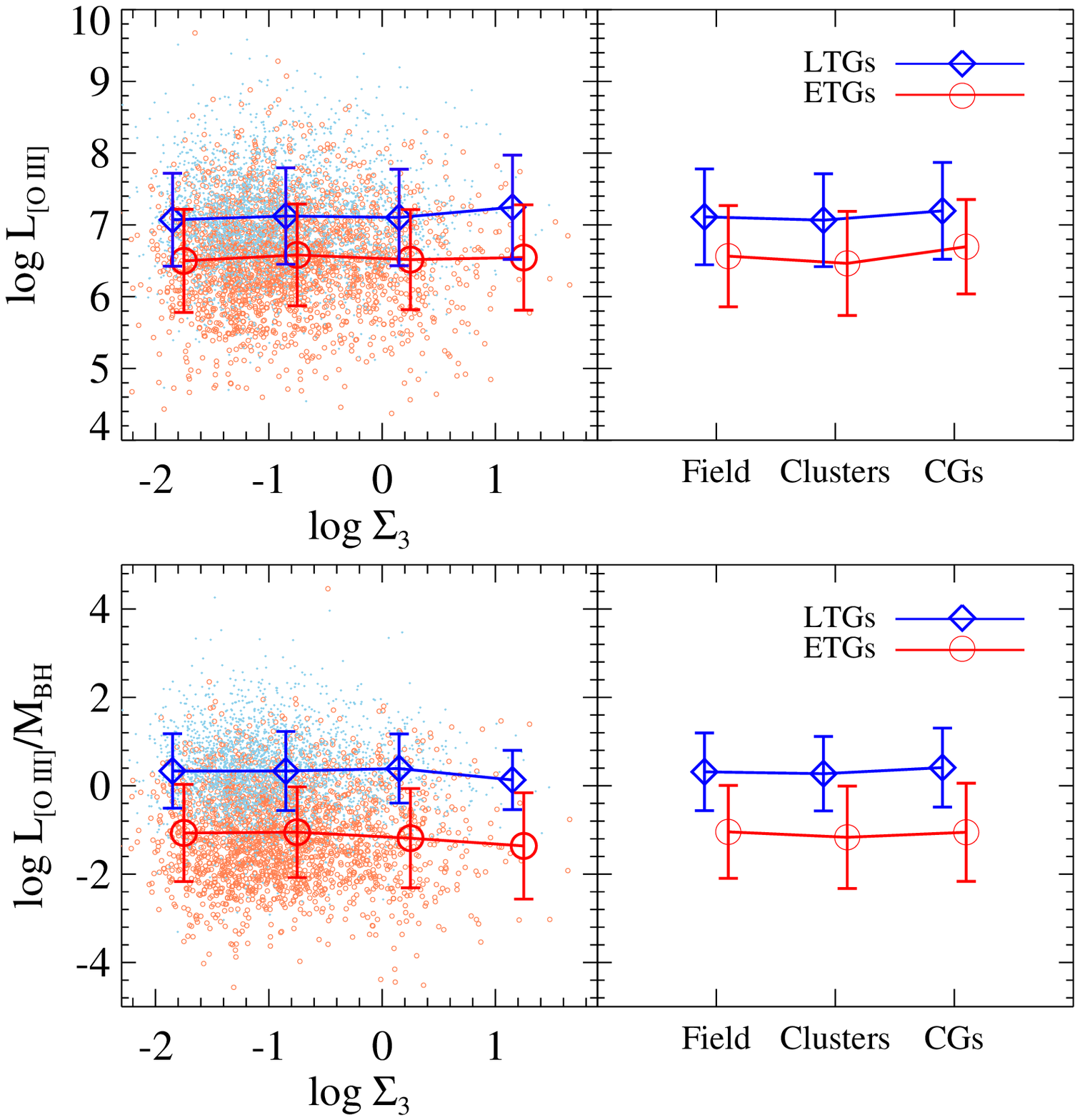}
\caption{Left panels: (a) $L_{[\rm O III]}$ vs. $\Sigma_{3}$ and (c) $L_{[\rm O III]}/M_{\rm BH}$ vs. $\Sigma_{3}$ of AGN in our total sample including field, clusters, and compact groups.
The small circles and diamonds show the distribution of early-type and late-type galaxies, respectively. 
The large circles and diamonds indicate the mean values of each type of galaxies, and the error bars mean standard deviations. 
Right panels: Mean values and their standard deviations (error bars) of (b) $L_{[\rm O III]}$ and (d) $L_{[\rm O III]}/M_{\rm BH}$ in three different galaxy environments. 
The definitions of symbols and error bars are same as left panels. 
Note that we arbitrarily shifted the large points to show symbols and error bars clearly in all four panels. }
\label{power}
\end{figure}

To further study the dependence of nuclear activity,
  we plot [O III] luminosities of AGN-host galaxies 
  (pAGN and composite galaxies from the BPT method) 
  as a function of $\Sigma_{3}$ (top left panel) and 
  of different environments (top right panel) in Figure \ref{power}.
It is apparent that 
  [O III] luminosities of late-type galaxies are, on average, 
  larger than those of early-type galaxies.
Interestingly,
  the [O III] luminosities for both early- and late-type galaxies
  do not show any environmental dependence.
  
In bottom panels, 
  we plot $L_{[\rm O III]}/M_{\rm BH}$ of AGN-host galaxies.
$L_{[\rm O III]}/M_{\rm BH}$ is proportional to the Eddington ratio,
  and can be used as an indicator of AGN power.
The $L_{[\rm O III]}/M_{\rm BH}$ again does not show
  any environmental dependence, similar to [O III] luminosities.
These results are consistent with 
  those in previous studies \citep{Hwa12a},
  suggesting that
  the triggering of nuclear activity depends on the environment,
  but the AGN power is not controlled by the environment.
    
\section{DISCUSSION}
\subsection{AGN fraction in compact groups: comparison with other studies}\label{compother}

In this section, 
  we compare the AGN fractions estimated in this study 
  with those for other compact group samples in previous studies.
Previous studies on the AGN fraction were mainly based on
  Hickson compact group galaxies \citep{Coz98,Shi00,Coz04,Mar10}.
For example, \citet{Coz98} first found that
  the AGN fraction for 82 bright galaxies in 17 Hickson compact groups
  is $\sim 40\%$ based on the BPT method.
Similar values of AGN fractions 
  (from 28\% in \citealt{Coz04} to 44\% in \citealt{Shi00})
  were reported based on similar sample sizes.
Recently, \citet{Mar10} studied the AGN fraction in Hickson compact groups 
  using the largest sample with 270 galaxies in 64 Hickson compact groups up to date. 
They obtained optical spectra for 200 galaxies,
  and compiled the spectra for the rest of sample galaxies in the literature.
According to their classification based on emission-line flux ratios,
  28.4\% and 14.4\% of their sample
  are classified as pure AGN and composite galaxies, respectively.
This corresponds to the AGN fraction (including composite galaxies) of 42.8\% 
  among the total sample. 
Although the AGN selection methods in these studies 
  are similar to the one in this study 
   (i.e., emission-line ratio diagrams),
   the previous values of AGN fractions seem higher than our fractions. 

\begin{figure}
\centering
\includegraphics[scale=0.9]{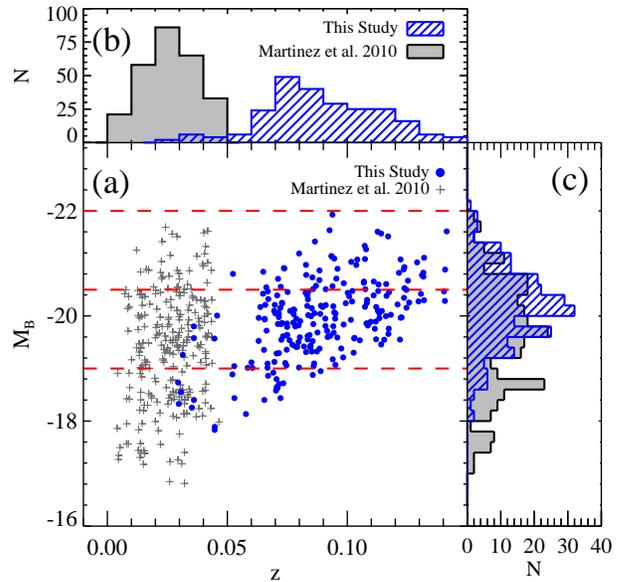}
\caption{(a) Absolute $B$-band magnitudes vs. redshifts for compact group galaxies in this study (circles) and in \citet[crosses]{Mar10}.
Redshift and magnitude histograms for these samples are shown in (b) and (c), respectively.
Hatched histograms represent the distribution of the galaxies in this study, while filled histograms denote that of compact group galaxies in \citet{Mar10}. 
Horizontal dashed lines denote the lines of demarcation of two magnitude bins that are used in Figure \ref{cg}.}
\label{redmb}
\end{figure}

To examine what makes the AGN fraction different 
  depending on galaxy samples, 
  we compare our compact group galaxy sample 
  with the one in \citet{Mar10}.
Figure \ref{redmb} shows absolute {\it B}-band magnitudes of 
  our compact group galaxies (blue filled circles) as a function of redshift.
We compute the {\it B}-band absolute magnitudes 
  using the SDSS $ugriz$ photometric data with the K-correct routine of \citet{Bla07}.
Because our sample is from the SDSS spectroscopic sample,
  the increase of magnitude limit with redshift is apparent in the figure.
For comparison, we also plot the galaxies in \citet[gray crosses]{Mar10}.
Our compact group galaxies are in the redshift range $0.03 \leq z \leq 0.15$, 
  but the galaxies in \citet{Mar10} are at $z \leq 0.05$.
We divide the galaxies into two magnitude subsamples 
  (dashed lines) to reduce the magnitude effects.

\begin{figure}
\centering
\includegraphics[scale=0.9]{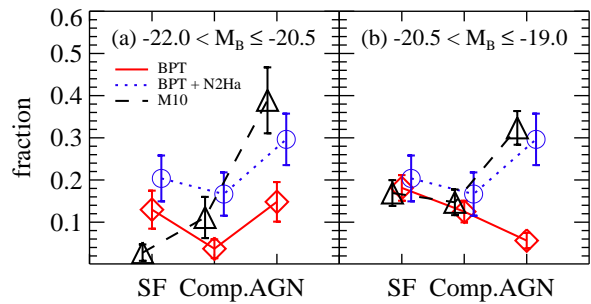}
\caption{Fractions of each spectral type for compact group galaxies in two magnitude bins; (a) $-22.0 < M_{B} \leq -20.5$  and (b) $-20.5 < M_{B} \leq -19.0	$.
Diamonds and circles, respectively, represent the fractions based on the BPT and the BPT+N2H$\alpha$ methods in this study.
Triangles represent those in \citet{Mar10}.
Each point is arbitrarily shifted along the x-axis to show symbols and error bars better.}
\label{cg}
\end{figure}

In Figure \ref{cg},
  we show the fraction of each spectral type 
  for two galaxy samples (this study and \citealt{Mar10}).
The left and right panels are for bright and faint galaxy subsamples.
In both panels, 
  the fractions of composite galaxies and pure AGN
  based on the BPT method in this study (red solid lines) 
  are significantly lower than those for \citet{Mar10}.
Note that \citet{Mar10} included weak emission-line galaxies,
  and used the N2H$\alpha$ method.
When we use the same classification method 
  as \citet{Mar10}  (i.e., N2H$\alpha$ method),
  the fractions in this study (blue dotted lines) are similar to those in 
  \citet[black dashed lines]{Mar10}\footnote{Because of increasing magnitude limit
   with redshift in our sample (see Figure \ref{redmb}), 
   the faint galaxy sample may not be complete.
   However, we confirm that the results do not change 
   even if we use the galaxies not affected by the magnitude limit.}.
This demonstrates that the estimated AGN fraction is strongly 
  affected by the AGN selection method.
This also emphasizes the importance of using a consistent AGN selection method 
  to have a fair comparison between different galaxy samples.

Table \ref{cg_agn} shows that
  the AGN fraction based on the BPT+N2H$\alpha$ method
  is higher than those based on other AGN selection methods.
This could result from a contamination 
  from `fake' AGN to the AGN sample in the N2H$\alpha$ method.
For example, 
  the galaxies with small H$\alpha$ equivalent widths (i.e., $< 3$ \AA) 
  are classified as `retired' or `passive' galaxies 
  in the WHAN method \citep{Cid11}.
However, these galaxies are classified as either
  star-forming, composite or AGN in the N2H$\alpha$ method 
  even though their emission lines are very weak.
Actually, 22 composite galaxies and 40 pure AGN 
  based on the N2H$\alpha$ method 
  have H$\alpha$ equivalent widths smaller than 3\AA,
  suggesting that the classification of these galaxies can be uncertain.

We also examine the mid-infrared colors of compact group galaxies 
  to identify AGN missed by the spectral diagnostics.
We use the Wide-Field Infrared Survey Explorer 
  ({\it WISE}, \citealp{Wri10}) all-sky source
   catalog\footnote{http://wise2.ipac.caltech.edu/docs/release/allsky/\#src$\_$cat},
   containing photometric data for over 563 million objects
   at four MIR bands (3.4, 4.6, 12, and 22 $\mu$m).
We use the point-source profile-fitting magnitudes,
  and restrict our analysis to the sources 
  with $S/N\geq3$ at 3.4, 4.6 and 12 $\mu$m bands.

\begin{figure}
\centering
\includegraphics[scale=0.9]{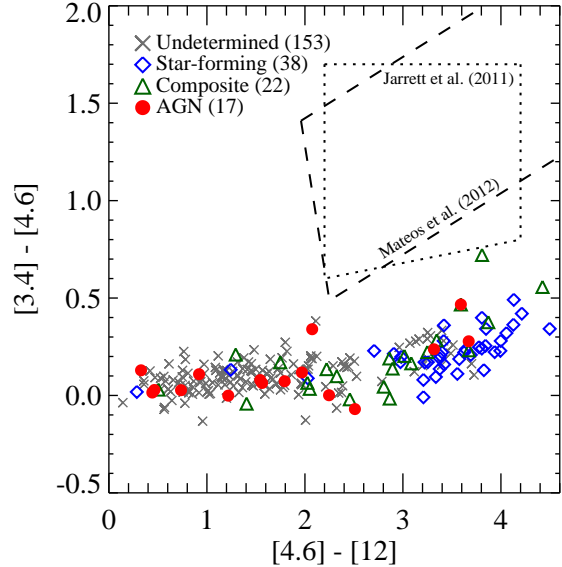}
\caption{Distribution of compact group galaxies in the {\it WISE} [3.4]-[4.6] color vs. [4.6]-[12] color diagram.
Different symbols indicate different spectral types based on the BPT method: star-forming (diamonds), composite (triangles), AGN (Seyferts + LINERs, circles), and undetermined (crosses).
Dotted and dashed lines are empirical criteria to select AGN suggested by \citet{Jar11} and \citet{Mat12}, respectively. }
\label{wise}
\end{figure}

The combination of {\it WISE} colors is
  useful for identifying dusty AGN-host galaxies 
  (e.g., \citealp{Jar11,Mat12,Hwa12b,Hwa12c}).
We show the {\it WISE} color-color distribution 
  of compact group galaxies in Figure \ref{wise}.
We also overplot two AGN selection criteria proposed by
 \citet[dotted line]{Jar11} and \citet[dashed line]{Mat12}.
Interestingly, none of compact group galaxies are 
  selected as AGN in this plot, 
  suggesting that there are few dusty AGN-host galaxies 
  in compact groups.
    
\subsection{Are the compact groups favored environments for the activity in galactic nuclei?}\label{compenv}

We find that the AGN fraction in compact groups is not the highest 
  among various galaxy environments.
If we assume that the activity in galactic nuclei is triggered 
  through galaxy-galaxy interactions, 
  the low AGN fraction in compact groups can suggest that 
  galaxy-galaxy interactions are not frequent in compact groups.
This is different from what is usually expected in compact group environment; 
  compact group galaxies would experience frequent interactions and mergers 
  because of their high galaxy number density and low velocity dispersion
  \citep{Rub91,For06,McI08}.
However, \citet{Zab98} pointed out that the galaxy interaction rate 
  in X-ray detected groups could be indeed 
  lower than for galaxy-dominated systems 
  because only 10$-$20\% of mass of X-ray detected groups is associated 
  with individual galaxies.

On the other hand, the compact groups are considered 
  to be dynamically old systems that already underwent interaction frequent stage.
The high fraction of early-type galaxies ($\sim65\%$) in the current epoch 
  also suggests that the compact group galaxies 
  could have experienced frequent galaxy interactions in early epoch.
For example, \citet{Wil09} found that the fraction of S0 galaxies 
  in $z\sim0.4$ groups is already much higher than those in the field,
  suggesting that the morphological transformation occurred at the early times.
The nuclear activity of compact group galaxies may also have been triggered 
  with morphological transformation at early epoch, 
  and then be turned off in the present epoch probably because of lack of fuel (i.e., gas).

In other words, the lower AGN fraction of early-type compact group galaxies than 
  for early-type field galaxies could be understood 
  in terms of lack of gas to fuel SMBHs.
Galaxy-galaxy interactions can produce a gas inflow 
  to the central regions of galaxies, 
  and then can trigger both nuclear activity and star formation (e.g., \citealp{Sto01}).
However, if there is no gas left in the interacting galaxies, 
  the nuclear activity cannot be triggered in spite of 
  frequent galaxy interactions \citep{Park09a,Park09b}.
As shown in Figures \ref{env} and \ref{sig}, 
  the AGN fraction of early-type galaxies is lower in high-density regions
  than in low-density regions.
This suggests that compact group galaxies contain less amount of gas 
  than field galaxies because the galaxies in compact groups might have
  consumed or lost their gas.
This argument is supported by the atomic gas depletion 
  observed in Hickson compact groups \citep{Ver01}. 
Verdes-Montenegro et al. found that the HI deficiency
  is stronger for compact groups with more early-type galaxies.
These imply that galaxy interactions involving early-type galaxies in compact groups
  do not trigger nuclear activity because of lack of gas. 
This interpretation is also valid for early-type cluster galaxies
  that might have consumed or lost most of their gas (see \citealp{Hwa12a}). 

For late-type galaxies, the AGN fraction varies little with the environment.
The atomic gas in late-type compact group galaxies also seems to be depleted 
  as in early-type compact group galaxies \citep{Ver01}.
However, the amount of molecular gas in late-type compact group galaxies 
  is still similar to that in field galaxies \citep{Mar12}. 
This suggests that the cold gas (mainly molecular gas) of late-type galaxies 
  in compact groups or clusters are not be stripped or consumed yet. 
This could be because late-type galaxies in compact groups or clusters
  are accreted recently \citep{Biv04,Hwa08}.
Similarly, \citet{Hai12} concluded that
  X-ray AGN in their galaxy clusters at 0.15$<z<$0.30
  appear to be an infalling population.
These results suggest
  that late-type AGN host galaxies in compact groups or clusters
  can still feed their SMBHs to be active
  using the gas that they keep from the field \citep{Mar09,Hag10,Hwa12a}.
Therefore, the AGN fractions for late-type galaxies are similar 
  in high- and low-density regions.

We also find that there are few dusty AGN-host galaxies 
  in compact groups (see Figure \ref{wise}).
Considering that the dust depletion in cluster galaxies 
  is strongly connected to the atomic gas depletion \citep{Cor12}, 
  the lack of dusty AGN in compact groups is consistent with expectation \citep{Ver01}.


\section{SUMMARY}
Using 238 spectroscopically selected member galaxies 
  in 58 compact groups at $0.03 \leq z \leq 0.15$,
  we study the activity in galactic nuclei in compact groups.
We also compare nuclear activity of compact group galaxies 
  with those of cluster and field galaxies.
Our primary results are summarized as follows.

\begin{enumerate}

\item Among the 238 compact group galaxies,
  we determine the spectral types of 83 galaxies with 
  strong emission lines based on the BPT method.
We also use WHAN and N2H$\alpha$ methods 
  to classify other 71 galaxies with weak emission lines.

\item We find a strong environmental dependence of 
  AGN fraction for early-type galaxies: 
  highest in the field, but lowest in cluster regions.
We confirm this trend with the surface galaxy number density, $\Sigma_{3}$.
These results suggest the nuclear activity in early-type galaxies
  is not strong in high-density regions because of lack of gas to fuel SMBHs.

\item The AGN fraction for late-type galaxies shows 
  little environmental dependence.
This can indicate that late-type galaxies in compact groups are accreted 
  recently, thus they can still keep their gas to fuel SMBHs.

\item $L_{[\rm O III]}$ and $L_{[\rm O III]}/M_{\rm BH}$ of AGN-host galaxies
   are higher in late-type galaxies than in early-type galaxies.
$L_{[\rm O III]}$ and $L_{[\rm O III]}/M_{\rm BH}$ of AGN-host galaxies
  do not show any environmental dependence. 
  
\item We find no dusty AGN-host galaxies in our galaxy sample of compact groups.
\end{enumerate}

\acknowledgments
We thank the anonymous referee for useful comments that improved the manuscript.
This work is supported in part by a Mid-career Researcher Program through an NRF grant funded by the MEST (No.2010-0013875).
J.S. is supported by Global Ph.D. Fellowship Program through an NRF funded by the MEST (No.2011-0007215).
H.S.H acknowledges the Smithsonian Institution for the support of his post-doctoral fellowship.
G.H.L. acknowledges support by the NRF Grant funded by the Korean Government (NRF-2012-Fostering Core Leaders of the Future Basic Science Program, No.2012-0002322). 
J.C.L. is a member of Dedicated Researchers for Extragalactic AstronoMy (DREAM) in Korea Astronomy and Space Science Institute (KASI).
This publication makes use of data products from the Wide-field Infrared Survey Explorer, 
which is a joint project of the University of California, Los Angeles, 
and the Jet Propulsion Laboratory/California Institute of Technology, 
funded by the National Aeronautics and Space Administration.
Funding for the SDSS and SDSS-II has been provided by the Alfred P. Sloan 
Foundation, the Participating Institutions, the National Science 
Foundation, the U.S. Department of Energy, the National Aeronautics and 
Space Administration, the Japanese Monbukagakusho, the Max Planck 
Society, and the Higher Education Funding Council for England. 
The SDSS Web Site is http://www.sdss.org/.
The SDSS is managed by the Astrophysical Research Consortium for the 
Participating Institutions. The Participating Institutions are the 
American Museum of Natural History, Astrophysical Institute Potsdam, 
University of Basel, Cambridge University, Case Western Reserve University, 
University of Chicago, Drexel University, Fermilab, the Institute for 
Advanced Study, the Japan Participation Group, Johns Hopkins University, 
the Joint Institute for Nuclear Astrophysics, the Kavli Institute for 
Particle Astrophysics and Cosmology, the Korean Scientist Group, the 
Chinese Academy of Sciences (LAMOST), Los Alamos National Laboratory, 
the Max-Planck-Institute for Astronomy (MPIA), the Max-Planck-Institute 
for Astrophysics (MPA), New Mexico State University, Ohio State University, 
University of Pittsburgh, University of Portsmouth, Princeton University,
the United States Naval Observatory, and the University of Washington. 


\clearpage


\end{document}